\documentclass[%
 aip,
 amsmath,amssymb,
 reprint,%
]{revtex4-1}

\usepackage{graphicx}
\usepackage{dcolumn}
\usepackage{bm}
\usepackage{color}
\usepackage[dvipsnames]{xcolor}
\usepackage[utf8]{inputenc}
\usepackage[T1]{fontenc}
\usepackage{mathptmx}
\usepackage{siunitx}

\begin{document}


\title[Ion Solvation in Dipolar Poisson models]{Ion Solvation in Dipolar Poisson models in a dual view}

\author{H\'elène Berthoumieux}
\author{Geoffrey Monet}
\affiliation{Sorbonne Universit{\'e}, CNRS, Laboratoire de Physique Th{\'e}orique de la Mati{\`e}re Condens{\'e}e (LPTMC, UMR 7600), F-75005 Paris, France}
\author{Ralf Blossey}
\email{ralf.blossey@univ-lille.fr.}
\affiliation{University of Lille, Unit\'e de Glycobiologie Structurale et Fonctionnelle (UGSF)\\ 
	CNRS UMR8576, 59000 Lille, France
}

\date{\today}

\begin{abstract}
We study the classic problem of ion solvation within the continuum theory of Dipolar-Poisson models. In this approach an ion is treated as a point charge within a 
sea of point dipoles. Both the standard Dipolar-Poisson model as well as the Dipolar-Poisson-Langevin model, which keeps the dipolar density fixed, are non-convex 
functionals of the scalar electrostatic potential $\phi$. Applying the Legendre transform approach introduced by A.C. Maggs [A.C. Maggs, Europhys. Lett. {\bf 98}, 16012 (2012)], 
the dual functionals of these models are derived and are given by convex vector-field functionals of the dielectric displacement and the polarization field. We show that the 
Dipolar-Poisson-Langevin functional generalizes the harmonic polarization functional used in the theory of Marcus for electron transfer rate to nonlinear regimes and can be quantitatively 
parametrized by molecular dynamics simulations for SPC/E-water. 
\end{abstract}
\maketitle
\maketitle

\section{Introduction} 
Modern soft matter electrostatics is commonly based on Poisson-Boltzmann (PB) theory \cite{dean14}. PB theory is a continuum description which starts from on a non-convex free energy 
functional $U$ of the electrostatic potential, $\phi$. The equilibrium properties of charged soft matter systems are evaluated from the Poisson-Boltzmann equation, the stationary point of $U$. 
The concavity of $U$, however, makes impossible a global minimization of the energy of the system when associated with other degrees of freedom such as, e.g., configurational energy. 
Maggs and collaborators have recently outlined a way to overcome this convexity problem \cite{maggs12,pujos14,blossey18}. The Legendre transform, a standard tool in statistical physics, 
allows to transform the non-convex PB theory into a convex vector-field based theory, in which the field variable is either, e.g., the dielectric displacement field ${\bf D}$ or the polarization field 
${\bf P}$. Although vectorial in character rather than scalar, such functionals enjoy an important range of applications. 

The most prominent example in the family of vector-field based electrostatic functionals is certainly the one used in the electron transfer theory pioneered by Marcus in the 1950's \cite{marcus56,bazant13,blumberger15}. 
This classical continuum model of electron transfer in solution relies on a harmonic polarization functional describing the dielectric state of the solvent and allowing for a linear response treatment to charge transfer 
\cite{marcus56} (for reviews, see \cite{bazant13,blumberger15}). The linear response/harmonic functional approximation for an aqueous solvent has been questioned by several authors \cite{zhou95,ichiye96,vuilleumier11} 
leading to a debate about the relevance of nonlinear effects. More recently, generalizations of Marcus theory have included molecular details of the solvent response \cite{jeanmairet19}. The authors of this work develop a more microscopic approach based on molecular density-functional theory, arguing in particular that ``Marcus theory does not take into account the molecular nature of the solvent which can break the 
linear assumption of solvent response. In such cases, we must resort to molecular simulation'' \cite{jeanmairet19}. We here advocate for an alternative approach based on continuum theory within dual electrostatics. We derive nonlinear functionals for the solvent and validate and parametrize them with molecular dynamics (MD) simulation. 

Phenomenological nonlocal functionals 
of polarization have been developed in recent years following a Landau-Ginzburg approach and were shown to capture the properties of water at the molecular scale \cite{kornyshev1997,maggs06,rottler2009,berthoumieux2015fluctuation,berthoumieux2018gaussian}.
While the qualitative behavior obtained with these models is in general agreement with the results of MD simulations, the amplitude of the structuration effects is overestimated in the linear models 
which raised doubts about the possibility to model water with pure electrostatic functionals \cite{fedorov2007}. Nonlinear terms modeling saturation effects have been therefore implemented also in a phenomenological manner \cite{kornyshevsat,berthoumieux2019}. However, these terms remain corrections and the corresponding functionals are still not sufficient to describe water subjected to the application of large external fields. 

In this work, we study two versions of generalized continuum models of water electrostatics, the Dipolar-Poisson model (D) and the Dipolar-Poisson-Langevin model  (DL) \cite{abrashkin07,levy12,levy13,frydel16}. 
Commonly, these models run under the name of Dipolar-Poisson-Boltzmann-(Langevin) models, but since our models disregard the equilibrated ions in bulk, we suppress the `Poisson-Boltzmann'-notions for clarity and brevity
in the following. For both these models, the free energy of the system is given in terms of the electrostatic potential $\phi$, obtained from a microscopic description of the solvent as an ensemble of point dipoles. Whereas the dipole density can vary in response to external field in the D-model, it is maintained constant for the DL-model. By performing the Legendre transform of the free energy density 
we exploit the duality between the scalar and vector field-based descriptions for these two models. We highlight the differences in their dielectric properties by considering the response of the fields to a point charge. 
We analytically determine the polarization functional for the DL model in the limits of low and high polarization and identify its range of validity. MD simulations for an SPC/E-water slab in the presence of an 
external field allow us to quantitatively parametrize the Dipolar-Langevin model. Consequently, we demonstrate that the Dipolar-Langevin model gives a reasonable, physically sound description of water despite its 
reductionist nature, opening a path that in the future may be useful for more advanced models.

\section{The Dipolar-Langevin model vs the Dipolar model}

As discussed in the introduction, our starting point are dipolar functionals and we will consider two cases: the first one obtained by considering an ensemble of dipoles of moment $p_0$ located on an incompressible grid of 
mesh \cite{abrashkin07} $a$, the Dipolar-Langevin model (DL)
\begin{equation}
\label{UDL}
 U_{\rm DL} = \int d^3{\bf r} \left[- \frac{\varepsilon}{2} (\nabla \phi)^2-
  \frac{1}{\beta a^3}\log \left(\frac{\sinh (\beta p_0  | \nabla \phi |)}{ \beta p_0  | \nabla \phi |}\right)\right]\,
\end{equation}
and the second one obtained by modeling water as a gas of dipoles, the Dipolar model (D),
\begin{equation}
\label{UD}
U_{\rm D}=\int d^3{\bf r} \left[- \frac{\varepsilon}{2} (\nabla \phi)^2-
\frac{1}{\beta a^3} \frac{\sinh (\beta p_0  | \nabla \phi |)}{ \beta p_0  | \nabla \phi |)}\right]+\frac{1}{\beta a^3}\,,
\end{equation}
where $\varepsilon$ is the vacuum permittivity and with $\phi$ as the electrostatic potential; $\beta = 1/k_B T$. Note that we have shifted $U_{\rm D}$ by the constant $1/\beta a^3$ such that the free energy 
vanishes for a vanishing potential.

We now introduce the electric field ${\bf E} = -\nabla \phi $ with a vector-valued Lagrange parameter ${\bf D}$, the dielectric displacement field, which yields
\begin{equation} \label{DLE}
U_{i} = \int d^3{\bf r} \left[-\frac{\varepsilon}{2}{\bf E}^2 - h_i({\bf E})  + {\bf D} \cdot (\nabla \phi - {\bf E}) \right]\, ,  
\end{equation}
where $h_i({\bf E})$ abbreviates the nonlinear expression in (\ref{UDL}), (\ref{UD})
with
\begin{eqnarray}
\label{hEDL}
h_{\rm DL}({\bf E})&=&\frac{1}{\beta a^3}\log \left(\frac{\sinh (\beta p_0  | E |)}{ \beta p_0  | E |}\right), \\
\label{hED}
h_{\rm D}({\bf E})&=&\frac{1}{\beta a^3}  \frac{\sinh(\beta p_0 | E |)}{\beta p_0 | E|}\, .
\end{eqnarray}
The free energy densities of both models, $u_i = -\varepsilon E^2/2-h_i(E)$, ($i$ = DL, D) are concave functions of the electrostatic field ${\bf E} = -\nabla \phi$.

\subsection{The Legendre transform of the Dipolar-Langevin and Dipolar models}

We introduce $\tilde{h}_i({\bf P})$ the Legendre transformation\cite{maggs12,pujos14} of $h({\bf E})$ defined as 
\begin{eqnarray}
\label{htilde}
\tilde{h}_i({\bf P}) &=& {\bf P}\cdot{\bf E} - h_i({\bf E}),\\
{\bf P}&=&\frac{d h_i({\bf E})}{d {\bf E}}
\label{PLegendre}
\end{eqnarray}
to express the free energy of the system as a functional of the conjugated field ${\bf P}$. It gives, for $ i = ({\rm D},{\rm DL}) $,
\begin{eqnarray} 
\label{PBP}
U_{i} & = & \int d^3{\bf r}  \Big[-\frac{\varepsilon}{2}{\bf E}^2  + \tilde{h}_i({\bf P}) - \phi\nabla \cdot {\bf D} + {\bf E} \cdot ({\bf D} - {\bf P})\Big]\, .
\end{eqnarray}
The variation of $U_i$ with respect to the electrostatic field ${\bf E}$ leads to the relation  ${\bf D}=\varepsilon{\bf E}+{\bf P}$ and thus identifies ${\bf P}$ as the polarization field. 
By replacing ${\bf E}$ by its mean-field value  $\varepsilon{\bf E}={\bf D}-{\bf P}$, we obtain the functional in the $P$-space:
\\
\begin{equation} \label{PB-DP}
\tilde{U}_{i} = \int d^3{\bf r} \left[\frac{({\bf D} - {\bf P})^2}{2\varepsilon} + \tilde{h}_i({\bf P})  - \phi \nabla \cdot {\bf D} \right]\, .
\end{equation}
 In the following, we consider only excitations ${\bf D}$ that satisfy the Gauss relation ($\nabla \cdot {\bf D} =0$ in the absence of free charge) and therefore drop the term $\sim \phi$ in the free energy density. 

The nonlinear Dipolar-Langevin (DL) and Dipolar (D) models can thus be studied in either $E$- and $P$-space and we now investigate their properties in these spaces.
We first compare the two $E$- and $P$-functionals with the corresponding linear continuum dielectric medium. To do so, we expand the dipole energy densities $h_i$ (i = D, DL), Eqs.(\ref{DLE})-(\ref{hED}), to
second order in ${\bf E}$, they are equal to  
\begin{equation} 
\label{hE}
h_{i,2}({\bf E}) = \frac{1}{2}\frac{ \beta p_0^2}{3 a^3} {\bf E}^2\, . 
\end{equation}
Using Eqs. (\ref{htilde},\ref{PLegendre}), one obtains in the $P$-space,
\begin{equation} \label{hP} 
\tilde{h}_{i, 2}({\bf P}) = \frac{1}{2}\frac{3 a^3}{ \beta p_0^2} {\bf P}^2\,.
\end{equation}
The harmonic approximations of the free energy densities in $E$-space, $f_i(E)=-\varepsilon{\bf E}^2/2-h_{i,2}({\bf E})$, and in $P$-space, $\tilde{f}_i(P)={\bf P}^2/2\varepsilon+\tilde{h}_{i,2}(P)$,
can be written as functions of the macroscopic response functions, -$\varepsilon\epsilon_w{\bf E}^2/2$, where $\epsilon_w$ is the relative dielectric permittivity and $P^2/2\varepsilon\chi$. 
The susceptibility $\chi$ controls the polarization response ${\bf P}_2$ of the linear dielectric medium to an external excitation ${\bf D}_0$, since ${\bf P}_2=\chi {\bf D}_0$.
By identification, one obtains the expressions 
\begin{equation}
\label{chiepsilon}
\epsilon_w \equiv 1+\frac{\beta p_0^2}{3\varepsilon a^3}\, ,\quad 1+\varepsilon\frac{3 a^3}{ \beta p_0^2} \equiv \frac{1}{ \chi}\, 
\end{equation}
as functions of the microscopic variables ($p_0$, $a$) of the D- and DL-models. In particular, the expression for $\chi$ makes the link with the phenomenological functional proposed by Marcus to study the electron 
transfer kinetic rates and the models studied here. We note particularly that the relations ($\ref{hE}$) and ($\ref{hP}$) are not only dual in the
field variables ${\bf E}$ and ${\bf P}$, but also in the model parameters, in the temperature dependence $\beta^{-1}$ and in $\varepsilon$.

\begin{figure*}
	\includegraphics{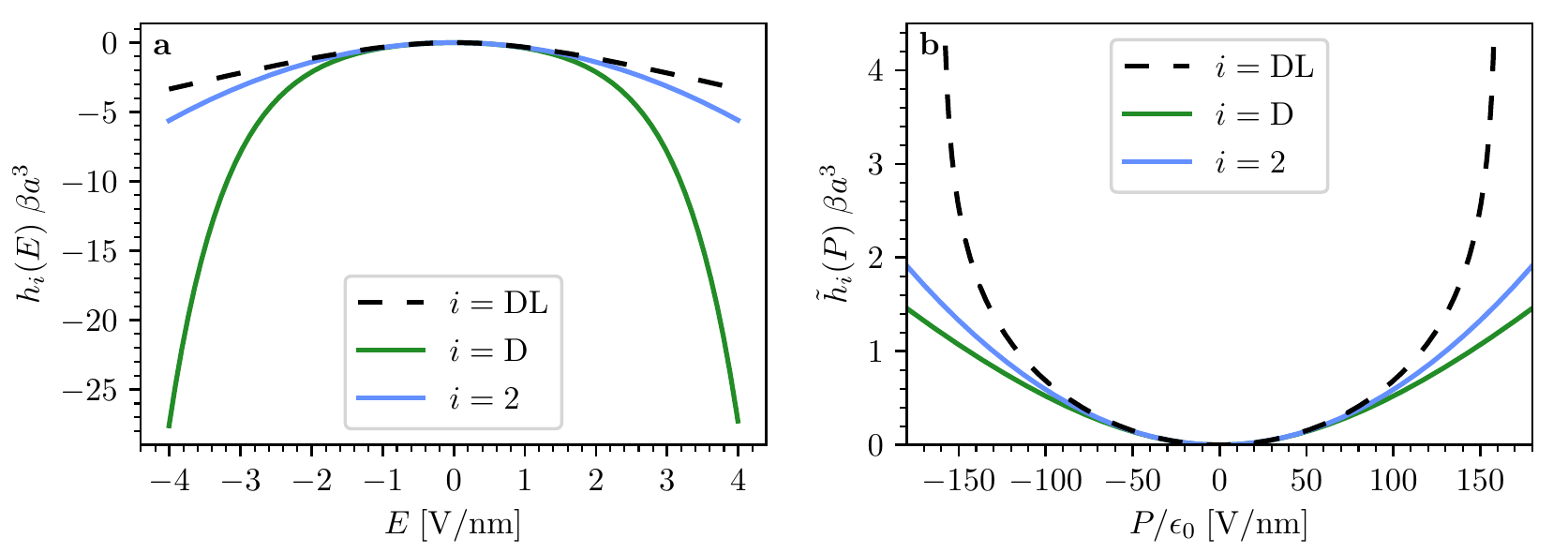}
	\caption{Free energy of the dipoles for the Dipolar and Dipolar Langevin models in $E$- ({\bf a}) and $P$-space ({\bf b}) . The dimensionless functions $h_{\rm D}({\bf P})\beta a^3$ and $h_{\rm DL}({\bf P})\beta a^3$  are 
	plotted using Eqs. (\ref{htilde}), (\ref{PLegendre}). The parameter values are $a = 0.17$ nm, $p_0 = 1.8$ D, $T = 300$ K.}
	\label{Fi1} 
\end{figure*}

In Fig.(\ref{Fi1}), the functions $-h_{\rm DL}({\bf E})$ (dashed black curve) and $-h_{\rm D}({\bf E})$ (green curve) and their harmonic approximation $-h_{{\rm DL},2}({\bf E})$ (blue curve) are plotted in panel {\bf a}.  
The corresponding functions for $P$-space are represented in panel {\bf b}.  As parameter set to reproduce the properties of water were used:  $p_0$ = 1.8 D corresponds to the dipole moment of one molecule and 
$a$ = 0.17 nm is adjusted to fix the relative dielectric permittivity $\epsilon_w$ to 78 corresponding to water. We note that it differs from the mesh size giving rise to the density of water which is given by $a_w$ = 0.3 nm. 
As one sees, for both the D- and DL-model the functions are concave in ${\bf E}$ and convex in ${\bf P}$.  We compare the value of the dipole energy density with respect to the harmonic approximation in  
$E$- and $P$- space. In the DL model, $\tilde{h}_{\rm DL}({\bf P})$ increases faster than the quadratic expansion in the $P$-space and saturates for large values of ${\bf P}$.
On the contrary, $h_{\rm DL}({\bf E})$ decreases more slowly than $h_{{\rm DL},2}({\bf E})$. We observe the opposite trend for the Dipolar model (green curve). 
A given excitation, ${\bf D}_0$, imposed on a medium described by the DL-functional will thus induce an under-response in $P$ and an over-response in $E$ compared to the corresponding linear medium. 
On the contrary, a medium described with the D-functional over-responds in $P$ and under-responds in $E$. 
 
These observations can be understood using the constitutive relation of electrostatics ${\bf D}=\varepsilon {\bf E} +{\bf P}$, which imposes that beyond the linear regime an over-response in $P$ or $E$ 
to an external field $D_0$ will be compensated by the saturation of $E$ or $P$. This duality is also a property of the Legendre transform, since the concave function of $x$, defined as
\begin{equation}
f(x)=-\frac{x^{2-\delta}}{2},\quad {\rm with} \quad \delta \leq  1
\end{equation}
grows more slowly in $x$ than the quadratic function $x^2/2$ for $\delta>0$, and faster for $\delta<0$.
Using the Legendre transformation, $f$ can be expressed as a function of the conjugated variable $y$,  
$f(x)=xy-g(y)$, which can be expanded to first order in $\delta$, giving rise to a $y-$powerlaw in $y^{2+\delta}$. 
A concave function that varies slower than a quadratic function is thus transformed into a convex function that increases faster than a quadratic function, as indeed observed in Fig.(\ref{Fi1}).


\begin{figure*}
	\includegraphics{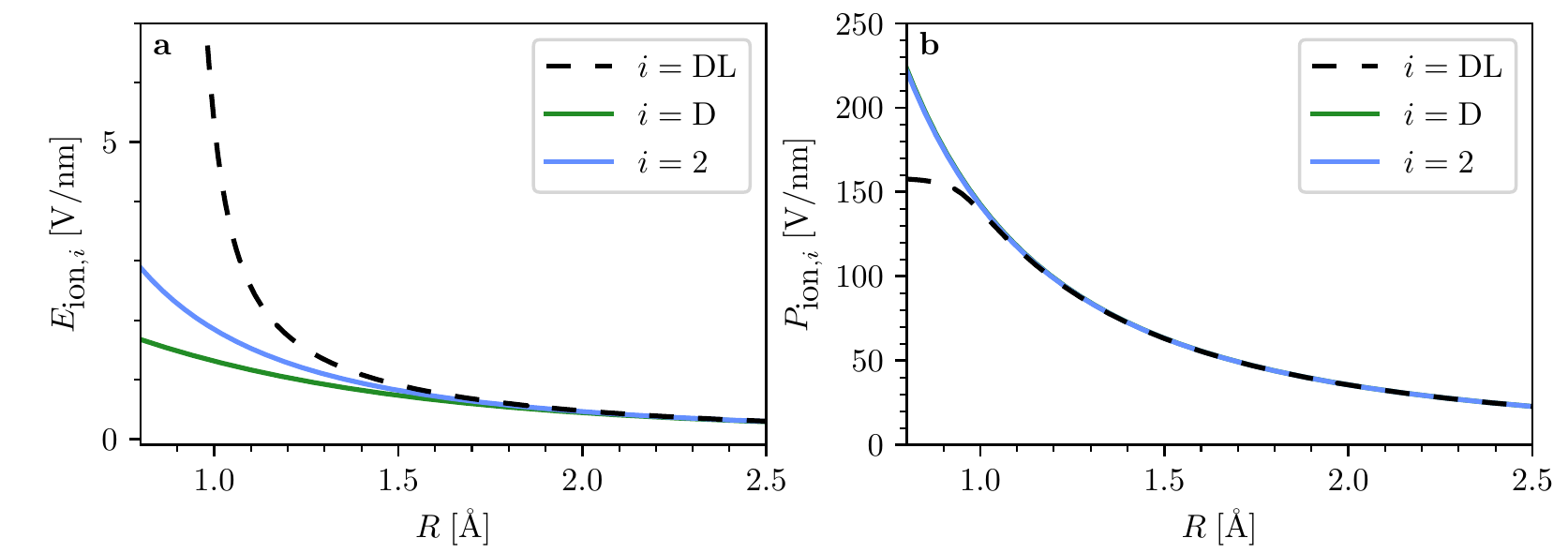}
	\caption{Electrostatic field and polarization response of a dipolar (D) and Dipolar Langevin (DL) medium to a monovalent ion.  $E_{\rm ion}$ ({\bf a.} )  and $P_{\rm ion}$  ({\bf b.} )  as a function of the distance to the punctual charge for D (green curve) and DL (dashed black curve) models obtained by solving Eq. (\ref{Eion}) and Eq. (\ref{Pion}). The blue curve corresponds to the response of the linear medium.  The parameter values are given in Fig.\ref{Fi1}.} 
	\label{F2} 
\end{figure*}

\subsection{Response to an ion in $P$- and $E$-space}
To deepen the comparison between the two models, in this subsection we derive numerically the four responses, in ${\bf E}$ and in ${\bf P}$ of the D- and DL-models to a point-charge distribution $e\delta(r)$ located in $r=0$, 
generating an excitation field 
\begin{equation}
{\bf D}_{0}(r)=\frac{e}{4 \pi r^2}{\bf e}_r\ .
\end{equation}
We determine the radial component $E_{\rm ion,i}(r)$ (with again (i = D, DL) thorughout) of the electrostatic response induced by the ion in the medium by minimizing the functional
\begin{equation}
\label{HEion}
U_{{\rm ion},i}=U_{i}+\int d^3r\, {\bf D}_0(r) {\bf E}
\end{equation}
with respect to ${\bf E}$. One obtains 
\begin{equation}
-\varepsilon E_{{\rm ion},i}-\frac{d h_i(E)}{d E} \bigg|_{E_{\rm ion,i}}= -D_0(r)  
\label{Eion}
\end{equation}
with $D_0(r)$ as the modulus of ${\bf D}_{0}(r)$.
Similarly, the radial component $P_{\rm ion,i}(r)$ of the polarization induced by the point charge is obtained by minimizing 
\begin{equation}
\label{HPion}
\tilde{U}_{{\rm ion},i}=\tilde{U}_{i}-\frac{1}{\varepsilon}\int d^3r\,{\bf D}_0(r) {\bf P }\,, 
\end{equation}
and solving the resulting equation
\begin{equation}
\label{Pion}
\frac{P_{\rm ion,i}}{\varepsilon}+\frac{d {\tilde h}_i(P)}{d P}\bigg|_{P_{\rm ion,i}}=\frac{D_0(r)}{\varepsilon}\, .
 \end{equation}
Eqs. (\ref{Eion}), (\ref{Pion}) are solved numerically for the Dipolar model (green curves) using Eq. (\ref{UD}) for $h$ and the Dipolar-Langevin model (dashed black curves) using Eq.(\ref{UDL}). 
The results are presented in Fig.(\ref{F2}). The linear electrostatic field $E_{\rm ion,2}(r)=e/4 \pi\varepsilon \epsilon_w r^2$ and polarization response  $P_{\rm ion,2}(r)=\chi e/4 \pi  r^2$ to the point charge are plotted in blue.

The plots confirm that a Dipolar-Langevin medium gives rise to an over-response in $E$ and under-response in $P$, in contrast to a Dipolar medium. One sees that the polarization response obtained using the 
DL-model is equal to the linear response at large distance of the point charge, which corresponds to a low excitation field, and then reaches a saturation value at small distances from the charge when submitted to a 
large excitation. The saturation in $P$-space is an essential property to describe a molecular polar solvent such as water. This saturation could be the origin of the decrease of the relative permittivity of aqueous electrolytes with  
increasing ion concentration. Indeed, each ion would create a sphere of only weakly responding water. 
The Dipolar-Langevin model, in contrast to the Dipolar model, reproduces this electrostatically `dead' shell in the vicinity of an ion. It also restores the linear behavior at large distances.

\section{Beyond the Marcus functional}

\subsection{Asymptotic regimes of low and high polarization}

Following our insights from the precious section, in this section we focus on the DL model in $P$-space. We derive expanded expressions of the free energy density in both limits of low and high polarization in order to obtain an analytical expression of the DL-polarization response to an ion over the full spatial range. For the sake of simplicity we drop the vectorial notations for the fields.
Using Eqs.(\ref{UDL}), (\ref{PLegendre}), the polarization as a function of the electrostatic field can be written as
\begin{eqnarray}
\label{PDL}
P(E) = \frac{ p_0}{a^3} \left({\rm coth} (\beta p_0 E)-\frac{1}{\beta p_0 E}\right)\, .
\end{eqnarray}
This expression can be expanded in the low-field regime, around $E=0$, or in the large-field regime for which the $\coth$-function in Eq.(\ref{PDL}) can be replaced by 1. The range of validity of the second regime can be 
estimated by considering $e^{-\beta p_0 |E|} = 0.01 e^{\beta p_0 |E|}$, {\it i.e.}, $E\geq1/2\beta p_0 \log(100)$ = 1.4 $10^9$V.m$^{-1}$ for water ($p_0$ = 1.8 D). 

We start with the low-field regime and expand Eq.(\ref{PDL}) around $E = 0$. As shown in Section II, the DL-model expanded to second order in $P$ can give rise to a quadratic dipole energy density given by Eq. (\ref{hP}). 
Two orders further - odd terms vanish - we get
\begin{equation}
\label{PDL4}
P_{\rm DL, 4}(E)=\varepsilon\left(\epsilon_w-1\right)E-\frac{p_0^4\beta^3}{45 a^3}E^3,
\end{equation}
a relation that can be inverted to obtain at the third order in $P$
\begin{equation}
\label{EDL4}
E_{\rm DL, 4}(P)=\frac{1}{\varepsilon(\epsilon_w-1)}P+\frac{9 a^3}{5 p^4_0\beta}P^3.
\end{equation}
Note that the index $n$ in the expressions ($P_n$, $E_n$) refers to the order of the expansion for the energy. 
We can thus write down the dipolar free energy density $\tilde{h}_{\rm{DL},4}(P)$ to fourth order in the polarization as
\begin{equation}
\label{hDL4}
\tilde{h}_{\rm{DL},4}(P) = \frac{1}{2\varepsilon}\frac{1}{\epsilon_w-1}P^2+\frac{9}{20 p_0\beta}\left(\frac{a^3}{p_0}\right)^3P^4.
\end{equation}
The correction in $P^4$ induces an increase of the free energy density for a given polarization $P$ compared to the linear model. It confirms that the DL-medium is associated with a higher free energy density than 
the linear medium for a given polarization and thus tends to saturate. 

In the high-field regime, the expression of $P$ as a function of $E$ is simply obtained by approximating  the  $\coth$-function by 1 in Eq.(\ref{PDL}) so that
\begin{equation}
\label{ET}
P_{\rm sat}(E) = \frac{p_0}{a^3}-\frac{1}{a^3 E \beta}, \quad E_{\rm sat}(P)=\frac{1}{\beta a^3(p_0/a^3-P)}.
\end{equation}
The polarization saturates to the value $p_{\rm sat}=p_0/a^3$, which is obtained when all the dipoles $p_0$ are aligned. In this situation,  the electrostatic field $E_{sat }(P)$ diverges. The dipole energy density in this regime $\tilde{h}_{\rm sat}(P)$  is written as
\begin{equation}
\label{hsat}
\tilde{h}_{\rm sat}(P) = \frac{1}{\beta a^3}\left(\log\left(\frac{2 p_0/a^3}{p_0/a^3-P}\right)-1\right).
\end{equation}
The linear and saturated regimes in response to an excitation are included in the Dipolar-Langevin model which could be a good candidate to model dielectric properties of water over a wide range of excitation amplitude. 
The question we address now is whether we can we propose a parametrization for the microscopic values $(p_0, a)$ that reproduces quantitatively these behaviors for simulated water.

\subsection{MD simulation-fit of the model}
 
\begin{figure*}
	\includegraphics{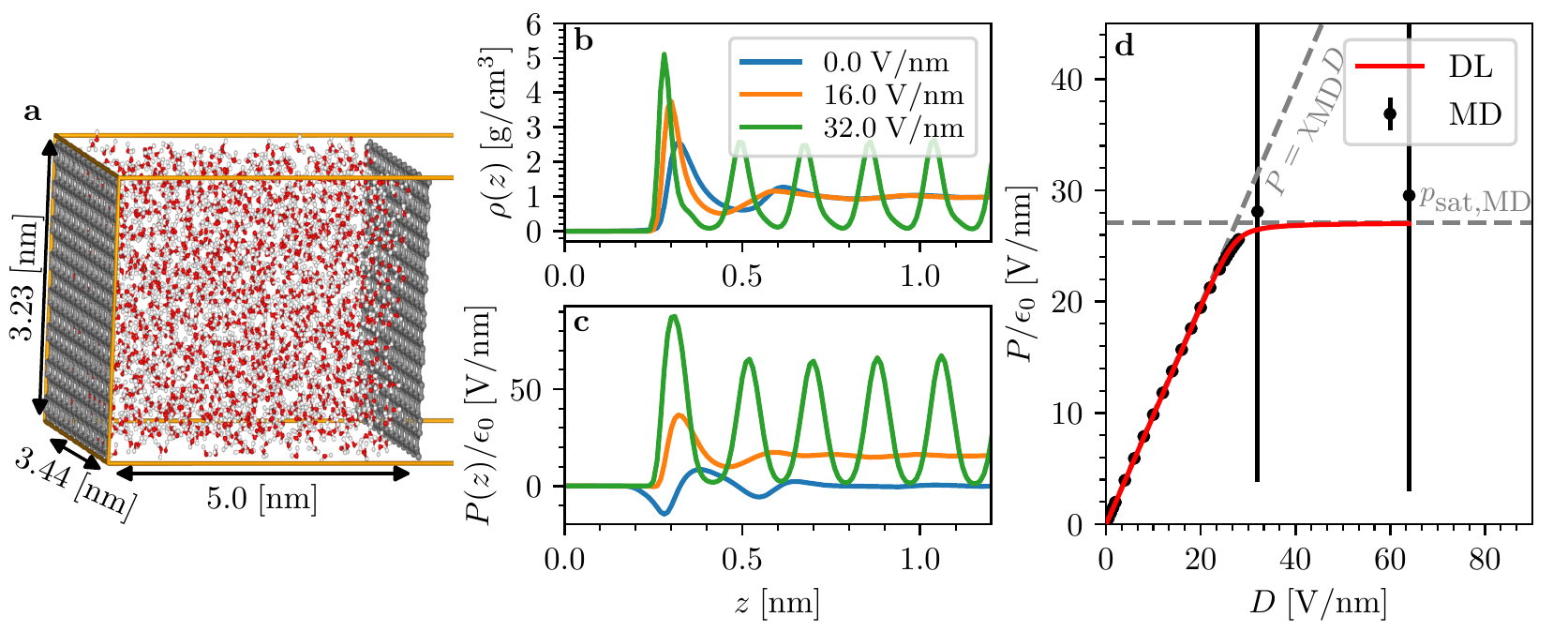}
	\caption{Molecular dynamics simulations of SPC/E water under excitation field. {\bf a.} Snapshot of the system composed of water molecules between two graphene sheets.  Mass density ({\bf b.}) and polarization ({\bf c.}) for three different excitation fields $D_0/\varepsilon=0$ V.nm$^{-1}$ (blue plot), $D_0/\varepsilon=16$ V.nm$^{-1}$ (yellow plot), $D_0/\varepsilon=32$ V.nm$^{-1}$ (green plot). {\bf d.} Polarization response as a function of the excitation field: MD simulations (black points); exact DL model (red plot); linear regime (dashed grey line), $P=\chi_{\rm MD}D_0$; saturation polarization  (horizontal dashed grey line) $P=p_{\rm sat}$. } 
	\label{F3} 
\end{figure*}

In order to enable an independent parametrization of the model, we performed molecular dynamics simulations of water confined between two graphene sheets in a slab geometry as illustrated by the simulation snapshot in 
Fig.(\ref{F3}) {\bf a}.
The slab walls are perpendicular to the $z$-direction and are made up of carbon atoms. The atoms are frozen, neutral and arranged in a hexagonal lattice. The two walls are separated by $L=\SI{5}{nm}$.
We use the SPC/E model for water which reproduces its dielectric properties best \cite{bonthuis12}.  It is associated with a relative permitttivity of 71. For the carbon-oxygen Lennard-Jones interaction we use the 
GROMOS53a6 force field. The simulation box is extended in the z-direction until it reaches a length equal to $3 L$. Thus, even if we consider a periodic system in the 3 directions of space, the periodized slab system is 
separated by a $2L$-thick void layer along the $z$-axis.

A $z$-directed excitation field ${\bf D}_{\rm MD, 0} = D_{\rm MD, 0} {\bf e}_z$ is applied between the two surfaces. After a short equilibration run of 0.01 ns with the $NVE$-ensemble, 
a long equilibration run of 2 ns was performed in the $NPT$-ensemble using the Berendsen thermostat $T=293.15$ K and pressostat ($P=10^{-3}$ katm). Finally, a simulation of 16 ns 
was performed in the $NVE$-ensemble.

We measure the response of the system for increasing amplitude of the excitation from 0 V.nm$^{-1}$ to 64 V.nm$^{-1}$.
The symmetry of the system imposes a $z$-directed response, ${\bf P}_{\rm MD}=P_{\rm MD}{\bf e}_z$, with
\begin{equation}
P_{\rm MD}(z) = \int_{-L/2}^{z} \rho_{\rm c, MD}(z)
\end{equation}
where $\rho_{\rm c, MD}(z)$ is the charge density of the fluid. In Fig.(\ref{F3}) {\bf b.}, 
the water mass density $\rho_{\rm MD}(z)$ (top panel) and the polarization, $P_{\rm MD}(z)$ (bottom panel) are plotted for different values of $D_{\rm MD, 0}$.  
For small excitations (up to 32V.nm$^{-1}$), $\rho_{\rm MD}(z)$  reaches
the bulk density and $P_{\rm MD}(z)$  a constant value on a distance of the range of the correlation length of water (1.5 nm). 
For large  $D_{\rm MD, 0}$, density and polarization oscillate with a period corresponding to the width of one molecule. 
We estimate the spatial mean of the polarization $P_m$ in the bulk, {\it i.e.} excluding the interfacial water of width $l_i$ = 1.5 nm \cite{monet2021}, as 
\begin{equation}
P_m = 1/(L-2l_i)\int_{-L/2+l_i}^{L/2-l_i} P_{\rm MD}(z) dz.
\end{equation}   
In Fig.(\ref{F3}) {\bf c.}, $P_m$ is plotted as a function of $D_0$ and one sees that the polarization is an affine function of $D_0$ at small values and reaches a plateau at large excitations. 
This highlights the linear and the saturated regime of the dielectric properties of SPC/E water. 

These two behaviors are captured by the DL-model and we fit the values of microscopic parameters $a$ and $p_0$ by measuring the slope $\chi_{\rm MD}$ of the linear part of $P_m(D_0)$ its saturation value $p_{\rm sat, MD}$.
By writing
\begin{equation}
	\chi_{\rm MD}=1+\varepsilon\frac{3 a^3}{ \beta p_0^2}, \quad p_{\rm sat, MD}=\frac{p_0}{a^3}\, ,
\end{equation}
we obtain $p_0$ = 9.62 D and $a$ = 0.51nm. By comparison, an SPC/E water molecule with $p_{\rm SPC/E} = 2.3$ D is associated with a mesh size $a_w=a_{\rm SPC/E}=$0.3 nm where $a^3$ is defined as the mean volume occupied by a water molecule.

\begin{figure}
	\includegraphics{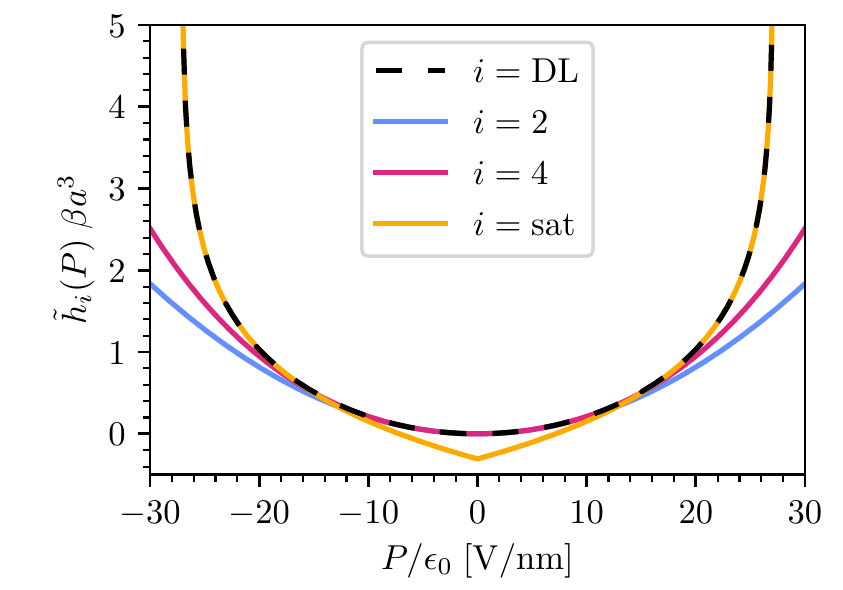}
	\caption{Dipolar free energy density in the low- and high-field regime for the DL-model. The fourth-order expression $\tilde{h}_{4,{\rm DL}}(P)$ (Eq. (\ref{hDL4}), pink curve)) and the large polarization expansion $\tilde{h}_{\rm sat,{\rm DL}}(P)$  (Eq. (\ref{hsat}), yellow curve) of the dipolar free energy density are compared to the exact density $\tilde{h}(P)$ (dashed black curve) and its harmonic expansion (blue curve). The cusp in $P=0$ for $\tilde{h}_{\rm sat,{\rm DL}}(P)$ is outside of the validity zone of the expansion and has no physical meaning. The parameters used are obtained from SPC/E water simulations: $p_0$ = 9.62 D, $a$ = 0.51 nm.} 
	\label{FDLh}
\end{figure}

The dipolar fluid described by Eq.(\ref{UDL}) reproducing the bulk permittivity and the saturation polarization of SPC/E water is composed of point dipoles with dipole moments 4 $\times$ larger and a density 7 $\times$ lower 
than SPC/E water. Using this parametrization, the asymptotic free energy densities $\tilde{h}_4(P)$ (low polarization, pink plot) and $\tilde{h}_{\rm sat}$ (high polarization, yellow plot) are plotted in Fig.(\ref{FDLh}). 
They are compared to the exact model $\tilde{h}_{DL}(P)$ (dashed black curve) and to the quadratic expansion $\tilde{h}_2(P)$ (blue curve), the Marcus functional.
The fourth-order expansion in the polarization and the high-field expansion allow to cover the whole range of field values with a very good precision. For the low-field regime, the fourth-order expansion brings a quantitative improvement when compared to the harmonic expansion that catches qualitatively the trend of the exact functional. For the high-field regimes, the polynomial expansions move away from  $\tilde{h}_{DL}(P)$ and the harmonic functional used by Marcus cannot be seen as a valid approximation anymore.

\subsection{Ion solvated in water}

Finally, we derive the polarization field around a point charge in the framework of the low- and high-field regime approximations by writing Eq.(\ref{Pion}) for $\tilde{h}(P)=\tilde{h}_{\rm DL,4}(P)$ and $\tilde{h}(P)=\tilde{h}_{\rm DL, sat}(P)$. We solve the two corresponding equations by taking advantage of the identity
\begin{equation}
	\frac{d {\tilde h}_i(P)}{d P}=E_i(P),
\end{equation}
and using the expressions of $E_i$ given in Eqs.(\ref{EDL4}), (\ref{ET}).

\begin{figure}
	\hspace{-0.9cm}
	\includegraphics{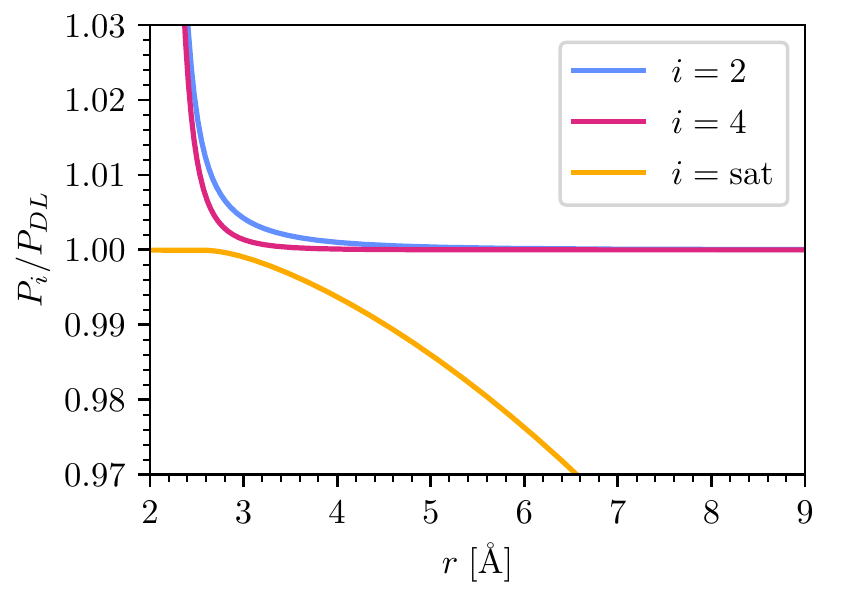}
	\caption{Rescaled polarization response to a monovalent ion calculated for low and high fields regimes of the DL model. The rescaled linear polarization $P_{\rm ion, 2}/P_{\rm ion, DL}$ (Eq. (\ref{Pion})) is plotted in blue, the rescaled fourth-order polarization $P_{\rm ion, 4}/P_{\rm ion, DL}$  is plotted in pink and the rescaled saturated polarization $P_{\rm ion, sat}/P_{\rm ion, DL}$ is plotted in yellow. The parameters used are given in Fig.(\ref{FDLh}).}
	\label{Fion}
\end{figure}

The polarization response $P_{\rm ion, 4}(r)$ obtained for the fourth-order expansion $\tilde{h}_{\rm DL,4}(P)$ is solution of a depressed cubic equation and is equal to
\begin{eqnarray}
	P_{\rm ion, 4}(r)
	&=&\frac{P_t}{2^{1/3}}\Bigg[\left(\frac{P_{\rm ion, 2}(r)}{P_t}+\sqrt{\frac{4}{27}+\frac{(P_{\rm ion,2}(r))^2}{P_t^2}}\right)^{1/3}\nonumber\\&+&\left(\frac{P_{\rm ion, 2}(r)}{P_t}-\sqrt{\frac{4}{27}+\frac{(P_{\rm ion,2}(r))^2}{P_t^2}}\right)^{1/3}\Bigg],
\end{eqnarray}
with the threshold polarization 
\begin{equation}
P_t=\frac{\sqrt{5}}{3}\sqrt{\frac{p_0^4\beta \epsilon_w}{a^3\epsilon_0(\epsilon_w-1)}}
\end{equation}
and the linear response $P_{\rm ion, 2}(r)=\chi D_0(r)$. 
The polarization in the solvation shell of the ion for low $r$ and high excitation is equal to
\begin{eqnarray}
& &P_{\rm ion, sat}(r)=\frac{p_0}{2 a^3}+\frac{D_0(r)}{2}\nonumber\\&-&\frac{\sqrt{\beta(a^3D_0(r)-p_0)^2+4a^3\varepsilon}}{2a^3\sqrt{\beta}}\, .
\end{eqnarray}
The polarization saturates towards $p_0/a^3$. 
In Fig.(\ref{Fion}) we plot the polarization as a function of the distance $r$ around an ion obtained for the linear (blue curve) and the two asymptotic regimes (4th order expansion in pink, saturation in yellow) rescaled by the 
exact polarization given by the DL-model. We use the parametrization determined in the previous section to model SPC/E water. As one sees, the linear model overestimates the response of the fluid for $r\leq$ 5 \AA. The fourth-order expansion brings a small gain in the 3-5 \AA range, but fails to reproduce the first solvation shell. The saturation model succeeds in this range (0.5-3 \AA) but has no meaning outside of this zone. Note that the range of validity of the different expansions depends on the charge of the ion. Roughly, the size of the saturation shell can be estimated as the radius for which the linear response $P_{\rm ion, 2}(r)$ equals the saturation polarization $p_{\rm sat}$ giving $r_{\rm sat}=Q^{1/2}\times(e a^3/4 \pi p_0)^{1/2}$, with $Q$ the valence of the ion. 
This shell possesses a 2.3~\AA~ radius  for a monovalent ion and 4~\AA~ radius for a trivalent ion.

\section{Discussion and Conclusions}

In this paper we have discussed the Dipolar-Langevin\cite{abrashkin07} and Dipolar\cite{levy12,levy13} Poisson models in the context of the recently formulated dual theory to the Poisson-Boltzmann approach
which leads to convex free energy functionals. The Dipolar-Langevin (DL) theory is derived from a microscopic model of water as dipoles fixed on a grid. The corresponding free energy density is a concave function 
of $\phi$ that we transform in a convex function of {\bf P}. We show that the DL-model in the $P$-space catches the linear polarization response of the medium at low excitation and its saturation in ${\bf P}$ on response to high excitation, as observed in MD simulations for various explicit models of water \cite{yeh1999}. We parametrize the DL model by fitting simulation measurements realized with the explicit SPC/E-water model, see Fig.(3) {\bf c}. 
When adequately parametrized, the DL model fits very well the nonlinear behavior of SPC/E water when submitted to a constant field $D_0$. Moreover, the Legendre transform of the analytical $\phi$-functional of the DL model (see Eq.(\ref{UDL})) combined to an expansion around $P = 0$ or around large polarization values gives access to an analytical expression of a polarization functional capturing the local properties of fluid, {\it i.e.} the bulk response function of a fluid to an excitation of any amplitude. We have thus obtained expressions of $P$ functionals that fit simulated water and are associated to a well-defined microscopic model of the system. The next step is to consider short-range correlated gas of dipoles as a microscopic input to try have access to nonlocal $P$-functionals.

Further, we have studied the solvation of a single ion in the DL model and derived analytic expressions for the polarization response in the two asymptotic regimes. The polarization is saturated in the first hydration shell of the 
ion of a radius that depends on the charge of the solute. This nonlinear effect is omitted by the harmonic functional used in the Marcus theory and could affect the derivation of the electron transfer rates. 
We furnish here an analytical framework to take this effect into account. 

In conclusion, we have demonstrated the usefulness of the dual approach to soft matter electrostatics. It can be easily extended beyond the specific model system we chose, the DL-model of point dipoles,
for more realistic continuum theories. Already within our simple models, our results on ion solvation demonstrate that a change of dual perspective can improve the quality of approximations. This is likely to 
be relevant in more complex situations, in particular in cases where non-electrostatic degrees of freedom have to be considered as well.
\\

{\bf Data availability statement.} The data that support the findings of this study are available from the corresponding author upon reasonable request.
\\

{\bf Funding.} HB and GM acknowledge funding under the Program Sorbonne Sciences - Emergences - 193256.
\\

\end{document}